\documentclass[fleqn,10pt]{elsart}
\usepackage{amssymb}
\usepackage{indentfirst}
\usepackage{epsfig}
\usepackage{epsf}
\usepackage{graphicx}

\newcommand{\D}{\displaystyle}

\journal{Nuclear Physics A}

\begin{document}

\begin{frontmatter}
\title{Parity-violating asymmetry of $W$ bosons produced in $p$-$p$ collisions}

\author[pku]{Xun Chen},
\author[pku]{Yajun Mao},
\author[ccast,pku]{Bo-Qiang Ma\corauthref{cor}}
\corauth[cor]{Corresponding author.} \ead{mabq@phy.pku.edu.cn}
\address[pku]{School of Physics, Peking University, Beijing 100871, China}
\address[ccast]{CCAST (World Laboratory), P.O.~Box 8730, Beijing
100080, China}

\begin{abstract}
The parity-violating asymmetry is an ideal tool to study the quark
helicity distribution in the proton. We study the parity-violating
asymmetry of $W^{\pm}$ bosons produced by longitudinally polarized
$p$-$p$ collision in RHIC, based on predictions of quark
distributions of the proton in the SU(6) quark-spectator-diquark
model and a perturbative QCD based counting rule analysis. We find
that the two models give nearly equal asymmetry for $W^+$ but that
for $W^-$ quite different. Therefore future experiments on such
quantity can help to clarify different predictions of the value
$\Delta d(x)/d(x)$ at $x \to 1$ in the proton.
\end{abstract}

\end{frontmatter}

\section{Introduction}
The helicity distribution is one of the three fundamental quark
distributions of the nucleon, and it has been explored for more
than two decades in deep inelastic scattering progresses. Deep
inelastic scattering (DIS) experiments of polarized electrons and
muons from polarized nucleons have confirmed us that the total
quark-plus-antiquark helicity sum is only about 1/4 to 1/3 of the
proton spin~\cite{disexp1,disexp2,disexp5,disexp3,disexp4}.
However, inclusive DIS experiments cannot provide information on
the polarized quark and antiquark density separately.
Semi-inclusive DIS measurements are one of the approaches to
achieve a separation of quark and antiquark helicity
densities~\cite{semidis1,semidis2}. But these experiments depend
on the details of the fragmentation process, so their accuracy is
limited. In these experiments, spin asymmetry is one of the most
important quantity to study the quark helicity distributions. The
Jefferson Lab Hall A Collaboration has released a precision result
on the neutron spin asymmetry~\cite{asyjoff1,asyjoff2} at large
$x$, which provides an examination on some known quark
distribution models. The disadvantage of semi-inclusive DIS
process can be avoided by using the production of $W$ bosons at
RHIC~\cite{prospectRHIC,hepph0405069}.

About ten years ago, Soffer and his collaborators pointed out that
the polarization of $u$, $d$, $\bar{u}$, and $\bar{d}$ in the
proton will be measured directly and precisely using the maximum
parity violation of $W$ bosons in $u\bar{d}\to W^+$ and
$d\bar{u}\to W^-$ at high $Q^2$~\cite{hepph9405250,b&s_npb445341}.
At leading order, the single spin asymmetry, or the
parity-violating asymmetry, of $W^{\pm}$ will approach to the
ratio $\Delta q(x)/q(x)$ when the absolute value of the rapidity
of $W$, $y_W$, is large. Some works have been done to calculate
this asymmetry with different quark distribution
models~\cite{craigie,b&s_plb314,hepph9405250,b&s_npb445341,Kamal_prd57,Gehrmann_NPB534,N&Y_NPB666,BS1993model,pddefine}.
The purpose of this work is to show that the parity-violating
asymmetry of $W^{-}$ bosons in longitudinally polarized $p$-$p$
process at RHIC can provide a crucial test of different
predictions on $\Delta d(x)/d(x)$ from a pQCD based analysis and
the SU(6) quark-spectator-diquark models, because it is very
sensitive to the value of $\Delta d(x)/d(x)$ at $x \to 1$ in the
proton.

The parity-violating asymmetry is defined as the difference of the
cross section of the left-handed ($-$) beams and that of the
right-handed ($+$) beams, divided by the sum of them, or using the
number density of the beams instead and normalized by the
polarization:
\begin{equation}
A_L = -\D\frac{\sigma_+-\sigma_-}{\sigma_++\sigma_-} =
-\D\frac{1}{P}\times \D\frac{N'_+-N'_-}{N'_++N'_-}.
\end{equation}

The production of $W$ is dominated by $u$, $d$, $\bar{u}$, and
$\bar{d}$ contributions, with some contamination of $s$, $c$,
$\bar{s}$, and $\bar{c}$ contributions, mostly through quark
mixing. Therefore $W$ production is an ideal tool to study the
spin-flavor structure of the nucleon.
The expected sensitivities at RHIC are also given in Ref.~\cite{prospectRHIC}.
P{\small HENIX} and S{\small TAR} both expect to get the enough $W^{+}(W^{-})$
data for analysis.

At RHIC, the $W^{\pm}$ will be produced in sub-processes
$u\bar{d}\to W^+$ and $d\bar{u}\to W^-$ by colliding beams of
protons spinning alternately left- and right-handed.

\begin{figure*}
\begin{center}
\resizebox{0.75\textwidth}{!}{
\includegraphics{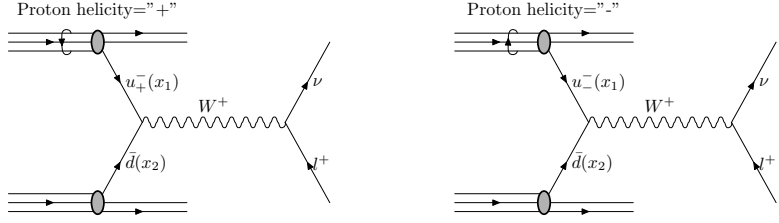}
}
\end{center}
\caption{A possible production of $W^+$ in $p$-$p$ collision, at
lowest order. The $u$ quark comes from the polarized proton and
the $\bar{d}$ quark comes from the unpolarized proton.}
\label{wproduction}
\end{figure*}

For Standard Model, the leading order production of $W^+$s, is
partly illustrated in Fig.\ref{wproduction}~\cite{prospectRHIC}.
Due to that the production of left-handed $W$ bosons violate
parity, the asymmetry can be constructed from polarized beams with
two directions . As Fig.~\ref{wproduction} shows, the $u$ quark
only comes from the polarized proton, the asymmetry should be
expressed as:
\begin{equation}
A_L^{W^+} = \frac{u^-_-(x_1)\bar{d}(x_2)-u^-_+(x_1)\bar{d}(x_2)}{u^-_-(x_1)\bar{d}(x_2)+u^-_+(x_1)\bar{d}(x_2)} =\frac{\Delta u(x_1)}{u(x_1)}.
\end{equation}

In the above equation, superscripts refer to the quark helicity
and subscripts refer to that of proton. $q(x_n)$ is short for
$q(x_n,Q^2)$, where $q=u,d$ and $n=1,2$, $Q^2=M_W^2$ and $M_W$ is
the mass of $W$ bosons.

For the situation that the $\bar{d}$ quark comes from the polarized proton, the asymmetry should be simply written as:
\begin{equation}
A_L^{W^+} = \frac{{\bar{d}}^+_-(x_1)u(x_2)-{\bar{d}}^+_+(x_1)u(x_2)}{{\bar{d}}^+_-(x_1)u(x_2)+{\bar{d}}^+_+(x_1)u(x_2)} = -\frac{\Delta \bar{d}(x_1)}{\bar{d}(x_1)} .
\end{equation}

In general, the asymmetry is a superposition of the two cases:
\begin{equation}
A^{W^+}_L = \frac{\Delta u(x_1)\bar{d}(x_2)-\Delta \bar{d}(x_1)u(x_2)}{u(x_1)\bar{d}(x_2)+ \bar{d}(x_1)u(x_2)} .
\end{equation}
The asymmetry of $W^-$ can be got easily by interchanging $u$ and
$d$:
\begin{equation}
A^{W^-}_L = \frac{\Delta d(x_1)\bar{u}(x_2)-\Delta \bar{u}(x_1)d(x_2)}{d(x_1)\bar{u}(x_2)+ \bar{u}(x_1)d(x_2)} .
\label{AWN}
\end{equation}
Higher-order corrections change the asymmetries only a little~\cite{Kamal_prd57,Gehrmann_NPB534}.

The kinematics of $W$ production and Drell-Yan production of
lepton pairs are the same. The momentum fraction carried by the
quarks and antiquarks, $x_n (n=1,2)$, can be determined by the
rapidity of $W$, $y_W$,
\begin{equation}
x_1=\frac{M_W}{\sqrt{s}}e^{y_W},\;\;\; x_2=\frac{M_W}{\sqrt{s}}e^{-y_W},
\end{equation}
where $\sqrt{s}$ is the center-of-mass energy. This picture is
valid only for the predominant production of $W$s at $p_T=0$. In
reality, the $W$ has $p_T$, which comes from the higher level
contributions such as $gu \to W^+d$ and $ud \to W^+g$, or the
primordial $p_T$ for the initial parton. But we must point out
that when $y_W \gg 1$, $A_L^{W^+} \approx \Delta u(x)/u(x)$, and
$A_L^{W^-} \approx \Delta d(x)/d(x)$. So if we take $p_T=0$ in
calculation, the results are also valuable at large $y_W$.

The experimental difficulty is that the $W$ is observed through
its leptonic decay $W\to l\nu$, and only the charged lepton is
observed. The connection of the rapidity of observed leptons and
that of $W$s has been described in detail in
Ref.~\cite{prospectRHIC}. And the expected cross section of $W$s
has also been estimated in that article.


\section{Quark Distributions}

\begin{figure}
\centering
\includegraphics[width=0.4\textwidth]{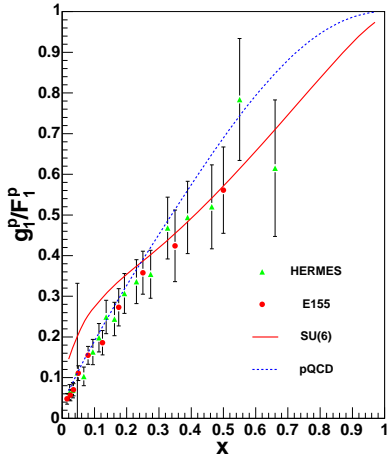}
\caption{Predictions on $g_1^p/F_1^p$ of quark-diquark model~\cite{diquark3} and pQCD based analysis~\cite{pqcd3} for $Q^2=4$GeV$^2$, and are compared to the experimental data from HERMES~\cite{disexp4} and E155~\cite{disexp5}. }
\label{g1p}
\end{figure}

\begin{figure}
\centering
\includegraphics[width=0.4\textwidth]{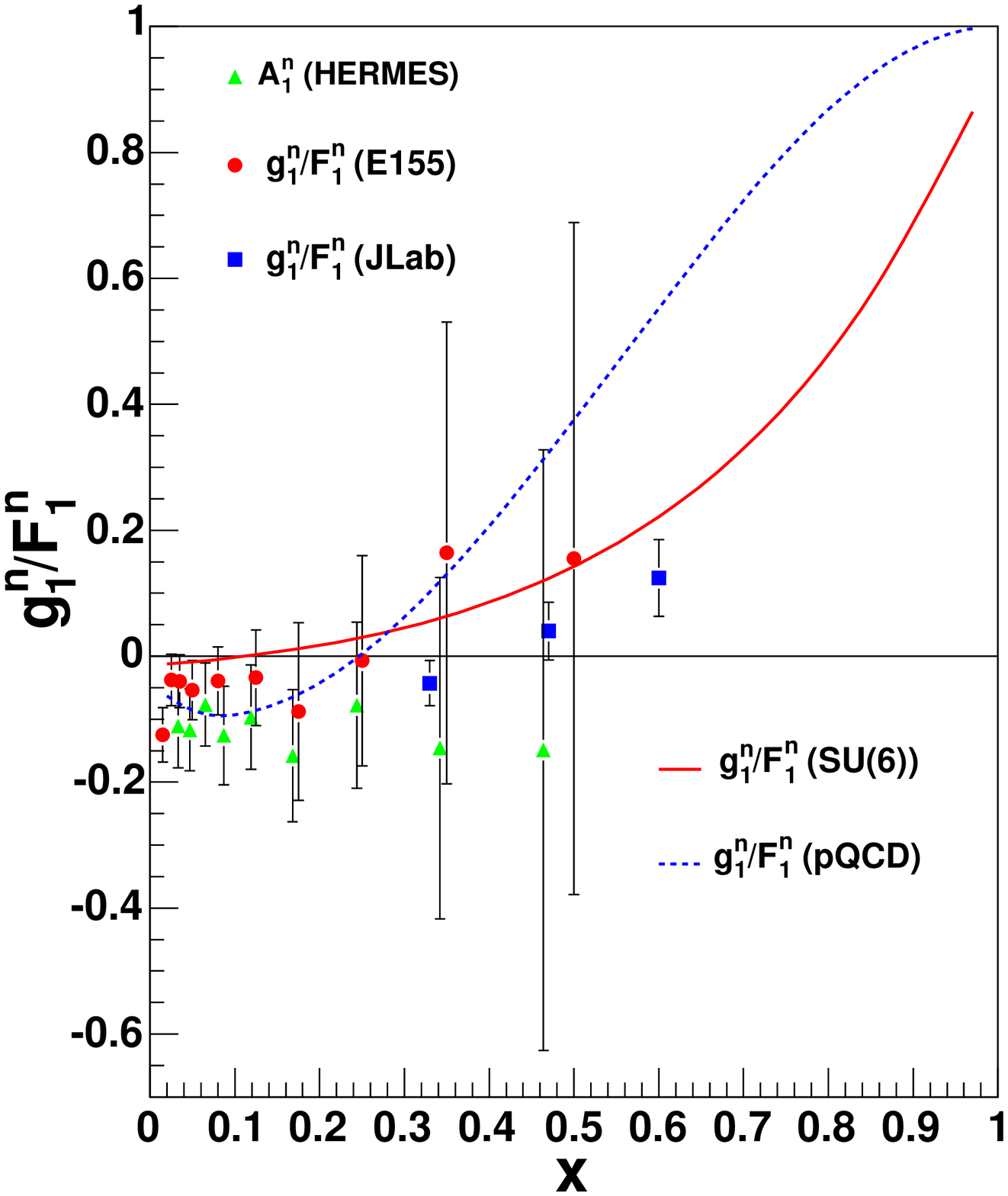}
\caption{Predictions on $g_1^n/F_1^n$ of quark-diquark
model~\cite{diquark3} and pQCD based analysis~\cite{pqcd3} for
$Q^2=4$~GeV$^2$, and are compared to experimental data from
JLab~\cite{asyjoff1,asyjoff2} and HERMES~\cite{disexp3}.}
\label{g1n}
\end{figure}

\begin{figure*}
\centering
\includegraphics[width=0.8\textwidth]{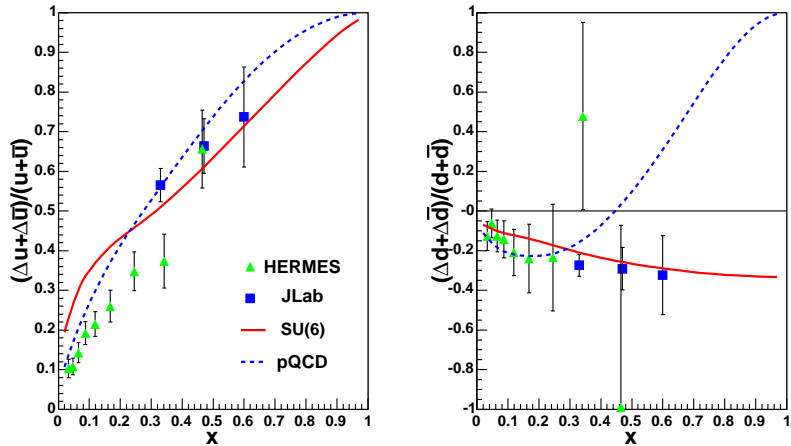}
\caption{Predictions on the flavor decomposition $(\Delta
u+\Delta\bar{u})/(u+\bar{u})$ and $(\Delta
d+\Delta\bar{d})/(d+\bar{d})$ of quark-diquark
model~\cite{diquark3} and pQCD based analysis~\cite{pqcd3} for
$Q^2=4$~GeV$^2$, and are compared to experimental data from
JLab~\cite{asyjoff1,asyjoff2} and HERMES~\cite{semidis2}.}
\label{duvsu}
\end{figure*}

From above equations, we can calculate the asymmetries of
$W^{\pm}$s if the quark helicity distributions are given. Two
known models, the quark-diquark
model~\cite{diquark1,diquark2,diquark3} and a pQCD based counting
rule analysis~\cite{pqcd1,pqcd2,pqcd3,pqcd4,pqcd5}, can provide
predictions of the quark distributions. Both models have their
advantages and have played important roles in the investigation of
various nucleon structure functions. However, there are still some
unknowns concerning the sea content of the nucleon and the large
$x$ behaviors of valence quark.  For example, there are still some
uncertainties concerning the flavor decomposition of the quark
helicity distributions at large $x$, especially for the less
dominant $d$ valence quark of the proton. There are two different
theoretical predictions of the ratio $\Delta d(x)/d(x)$ at $x\to
1$: the pQCD base counting rule analysis~\cite{pqcd3} predicts
$\Delta d(x)/d(x) \to 1$ whereas the SU(6) quark-diquark
model~\cite{diquark3} predicts $\Delta d(x)/d(x) \to -1/3$.
Experimental data of recent years show that both of these two
models are good in the predictions of $g_1^p$, and of the $\Delta
u(x)/u(x)$ extracted from it. But results of Jefferson Lab Hall A
Collaboration show that the pQCD prediction on $A_1^n$ deviates
from the data~\cite{asyjoff1,asyjoff2}. Figures
\ref{g1p},\ref{g1n} and \ref{duvsu} show the predictions of these
two models and the comparison with experimental data. It seems
that the pQCD predictions fit the data good at small $x<0.3$
whereas the SU(6) quark-diquark model works good at large $x>0.3$.


We can expect that the different predictions on $\Delta d(x)/d(x)$
of the two models may lead to prominent differences in calculating
the parity-violating asymmetry of $W^{-}$, due to the case that
$A_L^{W^-}\approx \Delta d(x)/d(x)$ when $y_W$ is large.

Detailed constructions of the quark helicity distributions in the
two models can be found in Ref.~\cite{consdetail}.

\begin{figure}
  \centering
  \includegraphics[width=.8\textwidth]{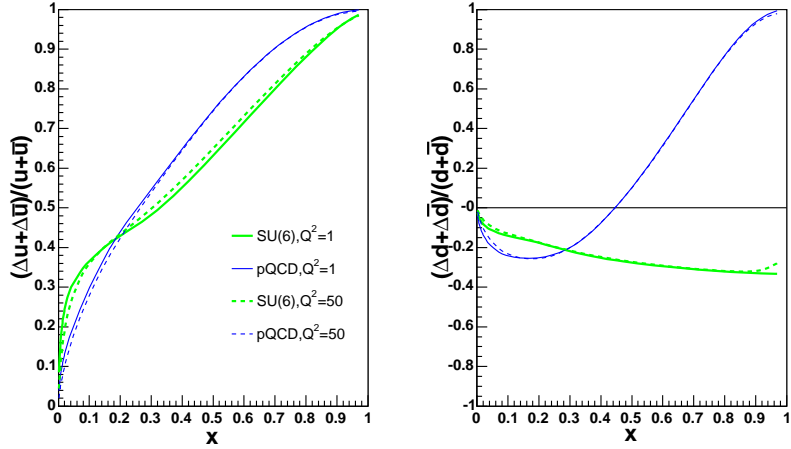}
  \caption{The $Q^2$ evolution of the helicity distributions.
  Prediction of the decomposition $(\Delta u+\Delta\bar{u})/(u+\bar{u})$ and $(\Delta d+\Delta\bar{d})/(d+\bar{d})$
   of quark-diquark model (thick curves) and pQCD based analysis (thin curves) for $Q^2=1$~GeV$^2$ (solid curves)
   and $50$~GeV$^2$ (dashed curves).}
  \label{fig:Cduvsu}
\end{figure}

To make realistic predictions of measurable quantities, we need
also to take into the sea quark contribution in the two model
constructions. In the quark-diquark case, this can be achieved by
adopting one set of unpolarized quark distribution parametrization
as input, and then use theoretical relations to connect the quark
helicity distributions with the unpolarized
distributions~\cite{diquark3,quarkdistribution}:
\begin{equation}
\begin{array}{c@{=}l}
\Delta u_v(x) & \D[u_v(x)-\frac{1}{2}d_v(x)]W_S(x)-\frac{1}{6}d_v(x)W_V(x),\\
\Delta d_v(x) & -\D\frac{1}{3}d_v(x)W_V(x).
\end{array}
\end{equation}
$W_S(x)$ and $W_V(x)$ are the Melosh-Wigner rotation
factors~\cite{Ma91} for spectator scalar and vector diquarks. We
use the valence quark momentum distributions $u_v(x)$ and $d_v(x)$
from quark distribution parametrization, but with $W_S(x)$ and
$W_V(x)$ from the model calculation~\cite{quarkdistribution}. In
this way we can take into account the sea contribution for the
unpolarized quark distribution from the input parametrization. The
energy scale dependence and the $Q^2$ evolution behaviors of the
quark distributions can be reflected by the explicitly $Q^2$
dependence of the input quark distribution parametrization. This
can provide a more reliable prediction for the magnitude and shape
of a physical quantity than directly from the model calculation.
CTEQ DIS~\cite{cteq5} (CTEQ5 set 2) quark distributions are used for
parametrization to investigate the evolution of the helicity distribution.
From Fig.~\ref{fig:Cduvsu}
we found that the ratios of $\frac{\Delta u(x)+\Delta\bar{u}(x)}{u(x)+\bar{u}(x)}$
and $\frac{\Delta d(x)+\Delta\bar{d}(x)}{d(x)+\bar{d}(x)}$
changed little with the increase of $Q^2$ in this model.

For the pQCD based analysis, we take the same consideration as the
above and make the following connection to relate the pQCD model
quark distributions with the parametrization:
\begin{eqnarray}
u_v^{\mbox{pQCD}}(x) &= &u_v^{\mbox{para}}(x), \\
d_v^{\mbox{pQCD}}(x)&=&\D\frac{d_v^{\mbox{th}}(x)}{u_v^{\mbox{th}}(x)}u_v^{\mbox{para}}(x),\\
\Delta u_v^{\mbox{pQCD}}(x)&=&\D\frac{\Delta u_v^{\mbox{th}}(x)}{u_v^{\mbox{th}}(x)}u_v^{\mbox{para}}(x),\\
\Delta d_v^{\mbox{pQCD}}(x)&=&\D\frac{\Delta
d_v^{\mbox{th}}(x)}{u_v^{\mbox{th}}(x)}u_v^{\mbox{para}}(x) ,
\end{eqnarray}
where the superscripts ``th'' means the pure theoretical
calculation in the pQCD analysis, and ``para'' means the input
from parametrization. The superscript ``pQCD'' means that the new
quark distributions keep exactly the same flavor and spin
structure as those in the pQCD analysis, but with detailed
$x$-dependent behaviors more close to the realistic situation.
This is equivalent to using a factor,
$u_v^{\mbox{para}}(x)/u_v^{\mbox{th}}$, to adjust each pure {\em
theoretically} calculated quantity to a more realistic pQCD {\em
model} quantity. In this way we can take into account the sea
contribution by using the sea quark distributions from
parametrization, while still keep the pQCD model behaviors of the
valence quark distributions.
We also use CTEQ DIS quark distribution functions as input to
investigate the $Q^2$ evolution of helicity distributions. The
results are drawn in Fig.~\ref{fig:Cduvsu}. We found the ratios
changed little with the increase of $Q^2$ in this model as those
in SU(6) quark-diquark model.

Thus we have two set of quark distributions of $\Delta q(x)$ and
$q(x)$, which keeps the same valence behaviors as in the quark
diquark model and pQCD based analysis prediction. In the SU(6)
model, the sea quarks are unpolarized, and only valence quarks are
polarized. Both of these two models do not give the polarized
distribution of the sea. HERMES data~\cite{expsea} show that the
absolute value of polarized sea quark distributions is much
smaller than the absolute value of polarized valence quark
distributions when $x>0.1$. So it does not matter for large $x$
and $y_W$ if we omit terms including $\Delta \bar{q}(x)$ in the
calculation.
The $Q^2$ evolution behavior of the decomposition indicates that
the asymmetry changes little when calculated with distribution
functions at different $Q^2$, thus the results and conclusion of
this work will not influenced qualitatively by the $Q^2$ evolution
effect.

\section{Numerical Calculations}
With the definition of $A_L$ and the quark distributions, we can
perform numerical calculations using different models.  For the
unpolarized quark distributions, we use the CTEQ LO
parametrization~\cite{cteq5} (CTEQ5 set 3) as input for both the
SU(6) quark-spectator-diquark model and pQCD based analysis
distribution functions described above. $Q=M_W=80.419$~GeV is
taken to get quark distributions from the CTEQ LO parametrization,
where $M_W$ is the mass of $W$ bosons.

\begin{figure*}
\begin{center}
\resizebox{0.75\textwidth}{!}{
\includegraphics{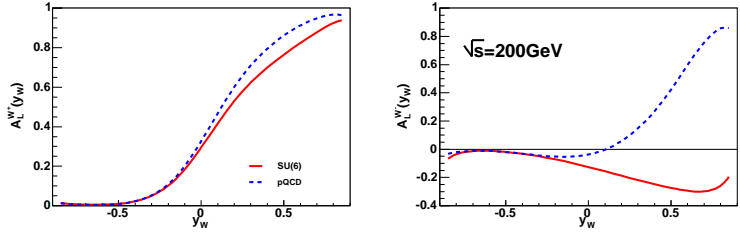}
}
\end{center}
\caption{The value of $A_L^{W}(y_W)$ in different models. The
solid curves correspond to the SU(6) quark-spectator-diquark
model, and the dashed curves corresponds to the pQCD based
analysis, for $\sqrt{s}=200$~GeV.} \label{y200}
\end{figure*}

\begin{figure*}
\begin{center}
\resizebox{0.8\textwidth}{!}{
  \includegraphics{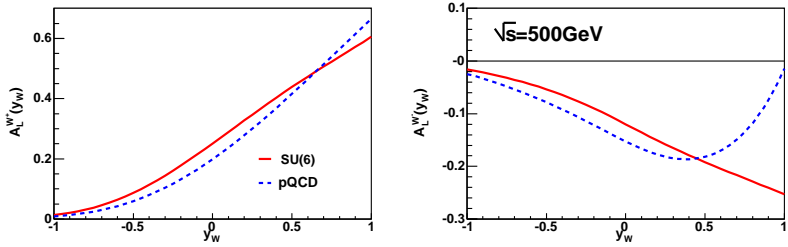}
}
\end{center}
\caption{The value of $A_L^{W}(y_W)$ in different models. The
solid curves correspond to the SU(6) quark-spectator-diquark
model, and the dashed curves correspond to the pQCD based
analysis, for $\sqrt{s}=500$~GeV.} \label{y500}
\end{figure*}

The polarized $p$-$p$ collisions at RHIC will take place at
center-of-mass energies of $\sqrt{s}=200\sim500$~GeV. So we
calculate $A_L^W$ at $\sqrt{s}=200$~GeV and $\sqrt{s}=500$~GeV.
Although the production of $W$ at 200 GeV is not feasible with the
designed luminosity at RHIC, we provide the calculation as a
reference to show the model dependent predictions at different
collision energies. Results of $A_L^{W^{\pm}}(y_W)$ are shown in
Fig.~\ref{y200} and Fig.~\ref{y500}.

It is clear that $A_L^{W^+}$ are nearly equal in these two models,
but $A_L^{W^-}$ show great difference between the two models at
both the two energy scales.

\begin{figure*}
\begin{center}
\resizebox{0.75\textwidth}{!}{
  \includegraphics{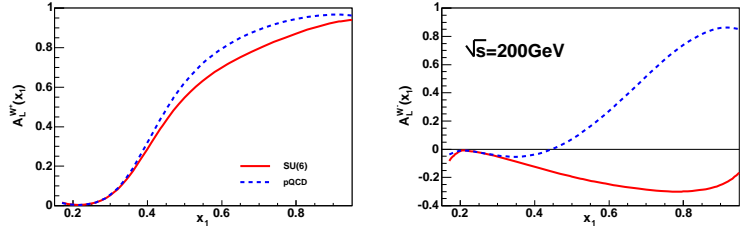}
}
\end{center}
\caption{The value of $A_L^W(x_1)$ in different models. The solid
curves correspond to the SU(6) quark-spectator-diquark model, and
the dashed curves correspond to the pQCD based analysis, for
$\sqrt{s}=200$~GeV.} \label{fig200}
\end{figure*}

\begin{figure*}
\begin{center}
\resizebox{0.75\textwidth}{!}{
\includegraphics{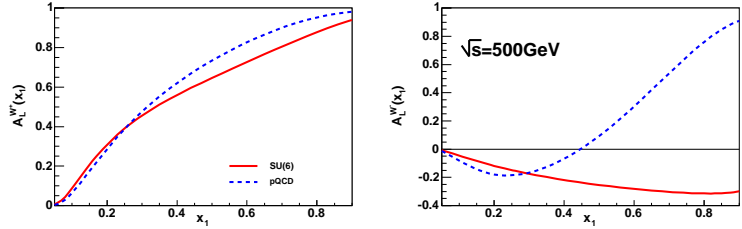}
}
\end{center}
\caption{The value of $A_L^W(x_1)$ in different models. The solid
curves correspond to the SU(6) quark-spectator-diquark model, and
the dashed curves correspond to the pQCD based analysis, for
$\sqrt{s}=500$~GeV.} \label{fig500}
\end{figure*}

As we mentioned before, $A_L^{W^-}\approx \Delta d(x)/d(x)$ when
$y_W$ is large, which means that $x_1$ is large. We may see the
difference more clearly by taking $x_1$ as the independent
variable. To examine this, we calculate $A_L^{W}$ again. Results
are shown in Fig.~\ref{fig200}, where $\sqrt{s}=200$~GeV, and
Fig.~\ref{fig500}, where $\sqrt{s}=500$~GeV.

The results convince us that the difference of $A_L^{W^-}$ in
these two models comes from the different predictions on $\Delta
d(x)/d(x)$ of the proton. We expect that experiments at RHIC can
give an examination.

\section{Summary}
The parity-violating asymmetry in the production of $W$ bosons in
longitudinally polarized $p$-$p$ process can be measured at RHIC.
It is sensitive to quark distributions of the proton. A pQCD based
analysis and the SU(6) quark-spectator-diquark model give obvious
different results on the parity-violating asymmetry of $W^-$,
which is sensitive to the value of $\Delta d(x)/d(x)$, a basic
quantity which is difficulty to be measured in other processes.
Therefore RHIC experiments results can give a powerful examination
on different predictions and enrich our knowledge on the spin
structure of the proton.

{\bf Acknowledgements}

This work is partially supported by National Natural Science
Foundation of China and by the Key Grant Project of Chinese
Ministry of Education (NO.~305001).


\end{document}